\documentclass[12pt,preprint]{aastex}
\begin{document}

\title{An annular gap acceleration model for $\gamma$-ray emission of
pulsars}

\author{G. J. Qiao$^{1}$, K. J. Lee$^{1}$, B. Zhang$^{2}$, H. G. Wang$^{3}$
\& R. X. Xu$^{1}$}

\affil{ $^1$Department of Astronomy, Peking University, Beijing
100871, China, gjn@pku.edu.cn \\
$^2$ Department of Physics, University of Nevada, Las Vegas,
Nevada 89154-4001, bzhang@physics.unlv.edu \\
$^3$ Center for astrophysics, Guangzhou University, Guangzhou
510400, China, cosmic008@263.net }

\begin{abstract}

{If the binding energy of the pulsar's surface is not so high
(the case of a neutron star), both the negative and positive charges
will flow out freely from the surface of the star. The annular 
free flow model for $\gamma$-ray emission of pulsars is suggested
in this paper.  It is emphasized that: (1). Two kinds of 
acceleration regions (annular and core) need to be taken into
account. The annular acceleration region is defined by the
magnetic field lines that cross the null charge surface within the light
cylinder. (2). If the potential drop in the annular region of a
pulsar is high enough (normally the cases of young pulsars), charges
in both the annular and the core regions could be accelerated and 
produce primary gamma-rays. Secondary pairs are generated in both 
regions and stream outwards to power the broadband radiations. (3). 
The potential drop in the annular region grows more rapidly
than that in the core region. The annular 
acceleration process is a key point to produce wide emission
beams as observed. (4). The advantages of both the polar cap and
outer gap models are retained in this model. The geometric properties of
the $\gamma$-ray emission from the annular flow is analogous to that
presented in a previous work by Qiao et al., which match the
observations well. (5). Since charges with different signs leave
the pulsar through the annular and the core regions, respectively, the
current closure problem can be partially solved.
}

\end{abstract}


\section{Introduction}\label{Sect1}

After more than thirty years of intense study, the origin of the
pulsed high energy emission from rotation powered pulsars is still
an unsolved problem. There are two types of radiation models, i.e.
the outer gap model  \citep{KMA85, KM85, CHR86a} and the magnetic
polar cap model. Within the polar cap models, two sub-types of model
exist. The vacuum gap model is based upon the assumption of strong
ion binding energy on the neutron/strange star surface \citep{RS75,
GS00, ZQLH97}. The space-charge-limited flow (SCLF) model
\citep{M74, FAS77, H81} assumes a low ion binding energy.
However in the previous polar cap models, only the core {cap}
has been considered, for a critical review see \cite{M82}. It is
suggested in this paper that two kinds of {accelerators}, in both the
annular {region} and the core {region}, must be taken into
account. The annular region is defined by the magnetic field lines
that cross the null charge surface(NCS). \cite{QLWXH04} emphasized
the importance of this annular region within the vacuum gap model.
The model is found suitable to reproduce the $\gamma$-ray pulse
profiles as well as the radio pulse profiles. In this paper, we
focus on the free flow {case}. Two major assumptions have been adopted
in this paper, which are the key points of the model. They are: 1). 
For an oblique rotator, particles in the magnetosphere flow out though 
the light cylinder and the returning currents are located in the 
annular region; 2). The plasma in the magnetosphere is not fully 
charge-separated.

In \S 2, we discuss our geometric setting and introduce the two
polar flow regions. Pair production and $\gamma$-ray emission are
discussed in \S 3, which focus on the properties of the annular
acceleration region. The conclusions are presented in \S 4 with some
discussions.

\section{Annular and core acceleration regions}

\begin{figure}
	\plotone{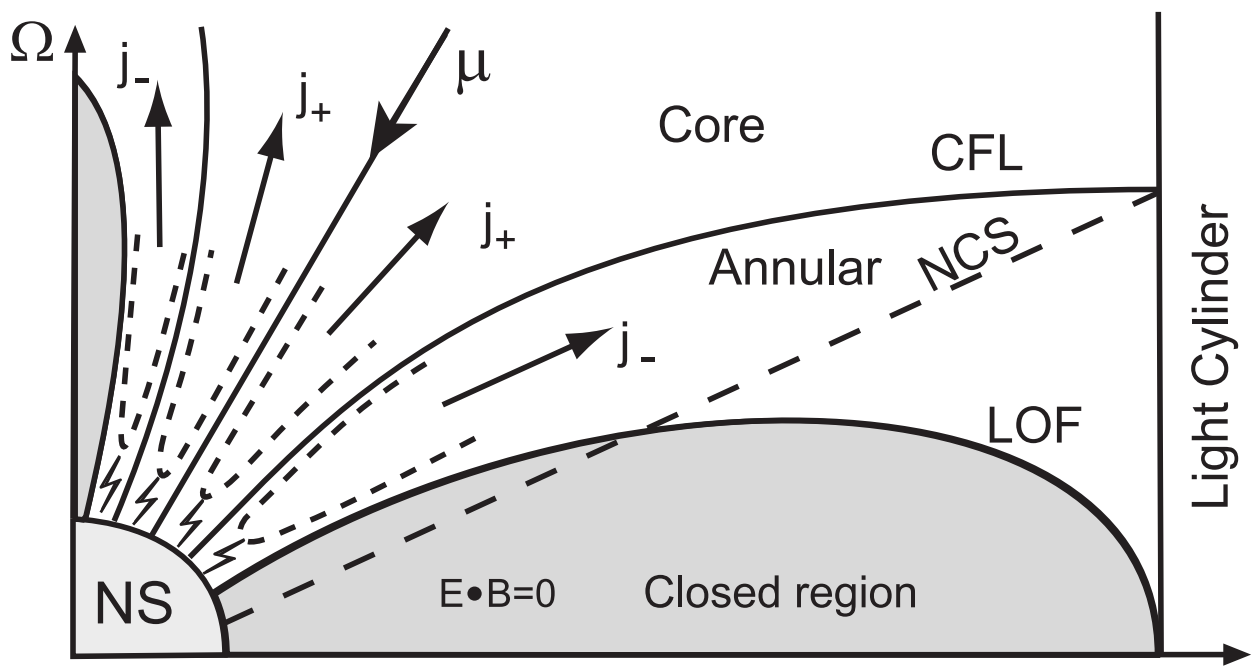}\caption {A schematic
diagram for the annular and the core regions for an inclined
rotator. $\Omega$ and $\mu$ are the rotational and magnetic axis,
respectively. The null change surface (NCS) is the surface where the
magnetic field is perpendicular to the rotation axis (i.e. ${\rm \bf
\Omega\cdot B}=0$). Line `CFL' and `LOF' are the critical field line
and the ``last" open field line respectively. Charged particles 
with opposite signs leave off the annular and the core cap regions,
respectively. The upper boundaries of the pair formation front are
denoted by the dashed lines. The ``flash'' symbols indicate the
locations with the maximal possibility to initiate the
pair-production cascade. } \label{global}
\end{figure}

The open field line region of the pulsar magnetosphere is divided
into two parts by the critical field lines. The part that contains
the magnetic axis is the core region, while the other part is the
annular region. The pulsar polar cap is also correspondingly divided
into the core and the annular polar cap regions (Fig.~\ref{global}).
For an aligned rotator, the radius from the magnetic pole to the edge of
the core cap is $r_{\rm cr}=(2/3)^{3/4}R (R \Omega/ c)^{1/2}$, while
the radius from the pole to the outer edge of the annular cap , i.e.
the radius of the whole polar cap, is $r_{\rm p}= R(\Omega
R/c)^{1/2}$ \citep{RS75}. Here $R$ and $\Omega$ are the radius and
the angular velocity of the star, respectively, and $c$ is the speed
of light.

For easy discussion in the following, throughout the text we will
focus on anti-parallel rotators, i.e. ${\rm\Omega \cdot \mu} < 0$.
The case of  ${\rm\Omega \cdot \mu} > 0$ could be derived by reversing
the signs, and all the conclusions in this paper remain valid.

For ${\bf \rm \Omega \cdot \mu} < 0$, the Goldreich-Julian charge
density ($\rho_{\rm{GJ}}$) {is positive in the region enclosed
by the null charge surface (NCS).}
For the core-region field lines, positive charges flow out through
the light cylinder. The supply of positive charges from the surface
compensate the deficit from the light cylinder. To maintain the
invariance of the total charge of the pulsar, one needs a current
that carries positive charges back to the star, or a current that
carries negative charges away from the star. For the {fully charge
separated magnetosphere, only one sign of charge presents at a given
location \citep{M79}.} In the annular region, the negative charged
particles at lower altitude can not pass the region with positive GJ
charge density at higher altitude. An outer gap then forms
beyond the NCS \citep{Hol73, KMA85, KM85, SMT01}.

However, resent pulsar magnetosphere simulations presented by some
authors \citep{S06, T06} give different results from that of
\citet{SMT01}. Furthermore, there is no a prior justification for a
fully charge-separated magnetosphere. Even if the magnetosphere is
initially charge-separated, the pair plasma generated from the outer
gap will soon fill the annular flux tube,
resulting in a quasi-neutral plasma. On the other hand, if one dismisses
the conjecture of a fully charge-separated magnetosphere, another
natural way to maintain the charge conservation of the pulsar is
simply by extracting negative charges directly from the annular polar
cap region (for the validity of the picture, see \S 3.3 for
details). In this paper, we explore such a possibility. It is found
that the negative charges stripped off from the stellar surface are
naturally accelerated in the annular region. This leads to a
particle acceleration model that keeps the geometrical advantages of
both the polar cap and the outer gap models, which is found suitable
to explain the $\gamma$-ray emission data (Dyks, Rudak 2003, Qiao
et al. 2004).

Since the positive and the negative charges are accelerated from the
core and the annular regions, respectively, the parallel electric
fields ($E_{\parallel}$) in the two regions are opposite, as has been
discussed by \cite{Stur71} and \cite{Hol75}. {As a result,
$E_{\parallel}$ vanishes at the boundary (i.e. critical field lines)
between the annular and the core regions. $E_{\parallel}$ also vanishes
along the closed field lines. Thus it is equal potential along the
closed field lines and the critical field lines. The potential on both
the closed field lines and the critical field lines should be equal to
the value at infinity.} We {also assume} $E_\parallel=0$ at the
star surface. When taking into account the $\gamma$-B process
for pair production, there exist two accelerators, one at the
annular region, and another at the core region. Furthermore, the pair
formation front also moves to further distances near the magnetic
pole.

One important issue is whether the secondary pairs can screen the
$E_\parallel$ developed in each region. In the core region,
since the primary charge density $\rho$ has the same sign as the
$\rho_{\rm GJ}$, the produced secondary pairs tend to get polarized
in the acceleration electric field and screen the field. This
usually happens especially if the primary $\gamma$-rays are produced
through curvature radiation \citep{HMZ02}.  For the annular flow, on
the other hand, since the primary charge density $\rho$ has the
opposite sign with respect to $\rho_{_{\rm GJ}}$, one has $\nabla^{2}
\Phi\propto (\rho-\rho_{\rm GJ})<0$. As a result, pairs can not
screen the acceleration electric field globally. Secondary pairs
will be accelerated in the residual electric field. As a result,
the charge acceleration region extends from the polar cap to higher
locations in the magnetosphere. This potentially matches the geometric
model proposed in Qiao et al. (2004). In the following, we will
elaborate the idea more quantitatively.

\section{Acceleration, pair production and $\gamma$-ray emission}

\subsection{Primary particle acceleration}
In flat space-time, the calculation involving the 1-D Poisson's
equation and the kinetic equation for charges to get the polar gap
potential drop was performed by \cite{M74}. In the Kerr space-time
{with small a Lense-Thirring angular velocity}, the Poisson's
equation is \citep{Bes90, MT92}.

\begin{equation}
\nabla\cdot(\frac{1}{\sqrt{\kappa}}\bigtriangledown\Phi)=-4
\pi(\rho-\rho_{\rm GJ}),
\end{equation}
where $\kappa=1-r_{\rm g}/r$, and $r_{\rm g}=2 G M/c^2$ is the
gravitational radius. To the lowest order approximation, one has
\begin{equation}
\label{rhogj} \rho_{\rm GJ}\simeq-\frac{({\rm \bf \Omega-{\rm \bf
\omega}_{\rm LT}}) \cdot {\rm \bf B}}{2 \sqrt{\kappa}\pi c},
\end{equation}
where $\omega_{\rm LT}=0.15 \Omega R^{3} r^{-3} I_{45}$ is the local
Lense-Thirring angular velocity, and $I_{45}$ is the moment of
inertia of the star in unit of $10^{45} \rm g \,cm^{2}$.

In order to reveal the qualitative difference between the annular
and the core cap regions (only within the primary accelerator), for
simplicity we only solve the 1-dimensional Poisson's equation
\begin{equation}
\frac{d}{ds}(\frac{1}{\sqrt{\kappa}}\frac{d \Phi}{ds})=-4
\pi(\rho-\rho_{\rm GJ})~. \label{1D}
\end{equation}
This equation is only valid when $s < r_{\rm P}$, but in
Fig.~\ref{gammah} a tighter limit is placed. A full 3-D treatment
is desirable to fully describe the electrodynamics of the system,
and we postpone it to a future work. {The 3-D calculation will
be reduced to 1-D result, when the solution is confined in a region
the transverse size of which is much larger than the longitudinal size,
i.e. $\nabla^2=\partial^2/\partial s_{\rm trans}+\partial^2/\partial
s_{\rm long}\simeq \partial^2/
\partial s_{\rm trans}$ for $s_{\rm trans}>>s_{\rm long}$.  
 Neglecting minor energy loss due to radiation, the
energy conservation law for particles in a given magnetic flux tube
is $\gamma m_{0}c^{2}+q \Phi=m_{0} c^{2}$, {where we have used the
condition $\Phi=0$ and $E_{\parallel}=0$ at the stellar surface. 
It should be
noted that here the condition $\Phi=0$ at the star surface can only be
used in the 1-D calculation and should not be regarded as the boundary
condition, since setting any $\Phi$ at the star surface do not change
the physical picture. The real physical boundary condition is
$E_{\parallel}=0$.}  This gives $m_0 c^2 d\gamma/ds=-q d\Phi/ds$. Here 
$\gamma$
is the Lorentz factor for the particles; $s$ is the distance along
the field line from the stellar surface; and $m_{0},q$ are the mass
and the charge of the particles we concern. In the annular region,
the primary particles are electrons, so $m_{0}$ and $q$ are the mass
and the charge of the electrons. In the core region, $m_0$ and $q$ are
the mass and charge of the ions pulled out from the surface. In 
curved space time in the pulsar vicinity, the current conservation
law can be expressed as $\sqrt{\kappa} J/B = \sqrt{\kappa}\rho v /B=
const$ (for flat space time case it is $\rho v/B=const$), where
we assume that the charged particles do not cross magnetic field lines.
Here $v$ is the velocity for the particles, and $J=\rho v$ is the
current density. The charge density at a given height can be then
expressed as $\rho = \sqrt\kappa_{0} v_{0} B \rho_{0} / \sqrt\kappa
v B_{0}$, where the subscript `0' denotes the values at the stellar
surface. Submitting the expression of $\rho$ and $\rho_{_{\rm GJ}}$
into eq.~\ref{1D}, we get
\begin{eqnarray}
\frac{d}{ds}(\frac{1}{\sqrt{\kappa}}\frac{d}{ds}\gamma)&=&\chi
\lambda^{-2} \left[\chi \frac{B v_{0} \sqrt{\kappa_{0}}}{B_{0} v
\sqrt{\kappa}}-\frac{\rho_{\rm GJ}}
{\rho_{\rm GJ 0}}\right ] \nonumber \\
\lambda^{-2}&=&\left |\frac{4 \pi q\rho_{\rm GJ 0}}{m_{\rm 0}
c^{2}}\right|, \label{gamma}
\end{eqnarray}
where $\lambda$ is the reduced Debye wave length for the surface
plasma, and a sign parameter $\chi$ is introduced. For ${\rm \bf
\Omega \cdot B} < 0$, one has $\chi=1, -1$ for the core and the
annular regions, respectively. For the $\rm \bf \Omega\cdot B>0$ case, 
the following calculations are also applicable. 
The boundary condition at the stellar
surface is $\rho = \rho_{_{\rm GJ}}$ in the core region, while $\rho
= -\rho_{_{\rm GJ}}$ in the annular region. The later is based on
the consideration that the negative polar cap current would
eventually compensate the GJ current loss at the light cylinder.

Equations (\ref{1D}) and (\ref{gamma}) are solved numerically for
both the core cap (CC) and the annular cap (AC) regions, as
presented in Fig.~\ref{gammah}. Analytical approximate solutions are
also derived. For the core cap, this is
$\gamma\simeq1+\sqrt{2}s/\lambda_{\rm core}$, while for the annular cap
it is $\gamma\simeq 1+s^{2}/\lambda_{\rm annular}^{2}$. We see that
while the particle Lorentz factor increases linearly in the core
region, it increases quadratically in the annular region, analogous
to the vacuum gap model \citep{RS75}. The analytical solutions are
also plotted in Fig.~\ref{gammah}, which show very consistent
results with the numerical solutions. Notice that in order to
compare the difference between the core and the annular regions, we have
assumed that the accelerated particles in the core region are
positrons. More realistic models involve positively charged ions.
This would change the Lorentz factor calculation (lower panel of
Fig.~\ref{gammah}) by a factor of mass ratio between the positron
and the ion.
\begin{figure}
	\plotone{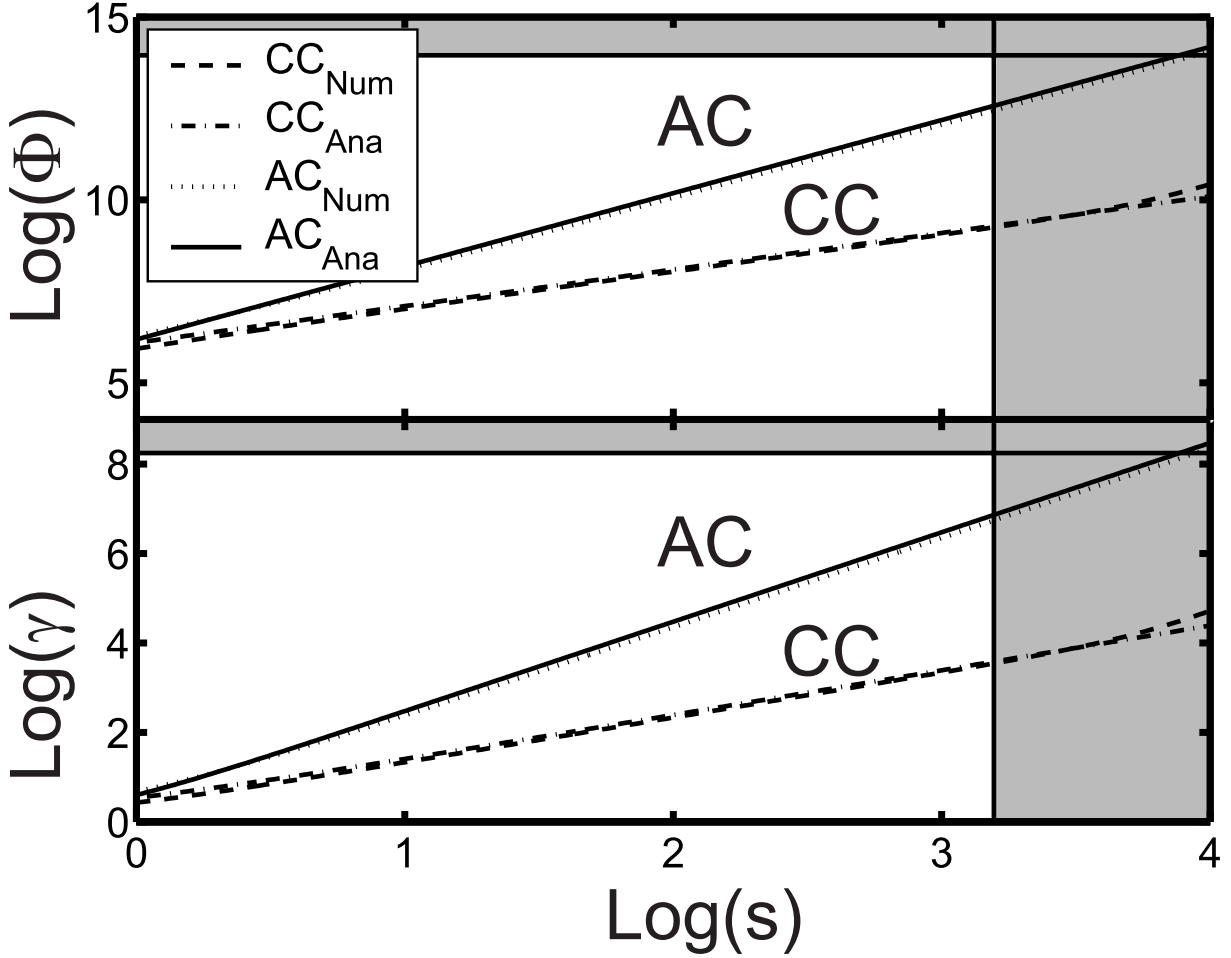}
\caption{The accelerated potential drop ($\Phi$) and Lorentz factor
($\gamma$) as a function of distance $s$ from stellar surface in the
AC and CC for an aligned anti-parallel rotator. $s$ is in units of
cm and $\Phi$ is in units of Volt. The legend ``AC'' and ``CC''
denote ``annular cap'' and ``core cap'', respectively. The
subscripts 'Num' and 'Ana' indicate the results from numerical
solution and analytic approximation, respectively. The shaded area
is the schematic regions where the 1-D annular polar cap
acceleration solution is no longer valid. The parameters used here are
common values for $\gamma$-ray pulsar (i.e. $p=0.1$ s, $B=10^{12}$
G).}\label{gammah}
\end{figure}

\subsection{Location of the pair formation front}

The primary particles gain very high Lorentz factors
($\simeq10^{3}\sim10^{5}$) within a short distance
($\simeq10^{2}\simeq10^{3}$ cm) (Fig.~\ref{gammah}). They will
radiate $\gamma$-ray photons via curvature radiation and inverse
Compton scattering \citep{RS75, ZQH97, ZQLH97, H81, ZHM00}. The
energetic $\gamma$-ray photons will be converted into electron and
positron pairs via the $\gamma$-B process. The condition to generate the
secondaries is $E_{\rm Mev}B_{\perp 12}/44\geq1/15$ if the magnetic
fields are not close to the critical values \citep{RS75}. For
curvature radiation, the typical $\gamma$-ray photon energy is
$E_{\rm cur}=\frac{2}{3}\frac{c}{\rho}\gamma^{3}\hbar$; for resonant
inverse Compton scattering, the typical photon energy is $E_{\rm
RICS}=2 \hbar \gamma \omega_{\rm B}$, where $\omega_{\rm
B}=\frac{eB}{m_{e} c}$ is the cyclotron frequency of electron
\citep{ZQH97}. The magnetic fields near the stellar surface may not
be pure dipolar \citep{RS75,GM02}. We then generically assume that
the curvature radius of the magnetic field lines is ${\cal R}$, so
that $B_{\perp}=B s/{\cal R}$, where $B$ is the total magnetic field
intensity. One can then estimate the typical length scale of pair
production. The typical height of pair formation front (or gap
height) in the core region has been estimated previously, e.g.
\cite{RS75,ZQH97,ZHM00,HMZ02}. Here we focus on the annular region.
In the annular cap, one has $\gamma\simeq1+s^{2}/\lambda^{2}\simeq
0.21 s^{2} B_{12} /P$, where $P$ is the pulsar period. Following the
method of \cite{ZHM00}, the height of the pair formation front in
the annular region can be estimated as
\begin{eqnarray}
\label{eqHeight} h_{\rm Cur}&\simeq& 4.2 \times 10^{3} (\rm cm)\,
B_{12}^{-4/7} P^{3/7}
{\cal R}_{6}^{2/7}\nonumber \\
h_{\rm RICS}&\simeq& 2.1\times 10^{4} (\rm cm)\, B_{12}^{-11/7}
P^{1/7} {\cal R}_{6}^{4/7}
\end{eqnarray}
where the subscripts 'Cur' and 'RICS' denote the cases of curvature
radiation, and resonant inverse Compton scattering, respectively. We
can see that for typical parameters, the pair formation length is in
the range of $10^{3} - 10^{4}$ cm. Since the solutions for
eqs.(\ref{1D}) and (\ref{gamma}) are valid only for acceleration
of primary particles, the solution breaks above the pair formation 
front. In Fig.~\ref{gammah} we use a shaded region on the right to 
denote the region above the pair formation front.

\subsection{Secondary pairs}

The behavior of the secondary pairs in the core region has been
widely discussed. The polarization of the pairs in the initial
electric field generates an electric field with the opposite
direction to compensate the initial electric field. In most cases,
the pair density is so high that the electric field is screened
\citep{HMZ02}.

The case of the annular region is different. Whether the particles
can be accelerated depends on the electric field direction. If the
electric field points towards the star in the annular region, the negative
charges will be accelerated outwards and positive charges will be
accelerated inwards. The only possibility for the electric field to
vanish along the magnetic tube is $\rho=\rho_{\rm GJ}$ associated
with trivial boundary condition. But the net free flow from the
stellar surface is negative, which is opposite to the local
$\rho_{_{\rm GJ}}$, thus the electric field can not be globally
screened, although it may be partially screened locally and temporarily.
The physical picture is that the negative charges move outward on a
inward-pointing electric field background (with positive charges
moving inward). The inward-pointing electric field is not a new idea.
This is in fact discovered very long time ago by Holloway (1973), Cheng
, Ruderman(1986), Michel(1979). They proposed the outer gap model,
in which the positive charges move inward, which indicates the
inward-pointing electric field or at least zero electric field.
Since our basic assumption is a quasi-neutral plasma, the negative
charges move outward, while positive charges move inward.

An interesting possibility within such a picture is that the annular
flow is non-stationary, as has been discussed earlier by
\cite{Stur71}. This is mainly caused by the interplay between the
un-screened electric field and the screening field of the pairs.
This may result in oscillating electric fields as envisaged by
\cite{LMJL05}, and may provide natural mechanisms to induce pulsar
inward emission \citep{D05a,D05b}.

\subsection{$\gamma$-ray luminosity}

The unscreened field extends at least to the NCS. Although with the
existence of pairs $E_\parallel$ is hard to model, electrons keep
accelerating in this extended annular region, and almost achieve the
maximum potential drop in the annular cap region, e.g.
\begin{equation}
\Phi_{\rm Max,Ann} = \frac{4 \pi^{2} R^{3} B}{c P^2}
\left[\left(1-\left (\frac{2}{3}\right
)^{3/4}\right)+\frac{\alpha}{2}\frac{1} {6^{3/4}}\right]^{2}.
\label{Phimax}
\end{equation}
For $\alpha=0$, it is reduced to the value of the aligned rotator
\citep{RS75}. The maximum particle luminosity from the annular
region is $\dot{E}_{\rm ann}=A \rho_{\rm GJ}c \Phi_{\rm Max,
Ann}\propto B^{2} P^{-4}(2+\alpha)$, where $A$ is the area of the
annular cap region. We can see that the annular cap power is
somewhat favorable to stars with large inclination angles. In
Fig.~\ref{gammah}, eq.(\ref{Phimax}) is placed as the upper limit of
$\Phi$ (denoted by the shaded region above).

An interesting inference from the above picture is that
$\gamma$-rays are preferably emitted in the annular flow. Qiao et
al. (2004) have investigated such a geometry model, and suggested
that the model could well fit the observed wide-beam $\gamma$-ray
pulse profiles as well as their radio pulse profiles. The
acceleration model proposed in this paper provides the physical
basis of that geometric model.

\section{Conclusion and discussion}

The annular free flow is taken into account in this paper. {In
the calculation, we assume that $E_{\parallel}=0$ at the star
surface and release the assumption that a pulsar magnetosphere must be
charge separated.} Several conclusions are derived. (1) There are
two separate polar cap regions that accelerate particles with
opposite signs; (2) Particle acceleration is more facilitated in the
annular cap region thanks to its quadratic growth of the electric
potential. It is then a favorable source of pair production. (3)
Secondary pairs can not screen the parallel electric field in the
annular region, so that electrons can be re-accelerated. The
acceleration flow may be unsteady. (4) The annular acceleration
region extends to the NCS or even beyond it. This leads to a fan-beam
$\gamma$-ray emission, which has been found suitable to interpret
the observed broad-band emission (Qiao et al. 2004). {(5) Both
the wide radiation beam observed in the high energy band and the current
closure problem for pulsar can be addressed in the same time, if 
$E_{\parallel}=0$ at star surface is assumed. }

{Whether the electric field is positive or negative in a given
region not only depends on the sign of charges in the region but
also the boundary conditions at the surface enclosing the region.
Here the electric field is solved in a 1-D model. The potential at the
star surface is arbitrarily chosen to be zero and the potential is
not a boundary condition. This is not true for 3-D calculations,
because the potential at the star surface and potential on the surface
of tube is used as the boundary condition. In the SCLF case,
boundary condition $\Phi=0$ is actually adopted at the star's
surface, the charges can be accelerated in the charge-depleted region.
However the boundary condition $E_\parallel=0$ is used at the star
surface in our case, so charge excess rather than charge depletion is
needed to accelerate particles in the core region. It should be noted 
that the Possion's equation is not well posed, if the electric field and
the potential are both specified as the boundary conditions at the same
places.}

The traditional outer gap model is based on the assumption of a
fully charge-separated magnetosphere. Our model is based on the
different assumption that negative charges are freely supplied from
the surface of the pulsar. This leads to a different acceleration
picture, in which the $\gamma$-ray emission region can be closer to the
null surface. In principle, our model suggests that a vacuum outer gap 
should not exist.

The re-acceleration and possible pair-production process beyond the
pair formation front involves more complicated physics and is
subject to further study. A more generalized 3-D treatment for both
the core and annular accelerators is desirable.

Both the core and the annular accelerators produce pairs that can emit
radio waves. The bi-drifting phenomenon (Champion et al. 2005) can be
well interpreted by assuming that the observed radio emission comes from 
both regions (Qiao et al. 2004b).

If pulsars are bare strange stars \citep{XQZ99}, vacuum gaps would
be formed. The separation between the core and the annular region in
that model has been discussed in Qiao et al. (2004).

\begin{acknowledgements}
We are very grateful to the referee for valuable comments, and 
Dr. Han, J. L. and Mr. Zhu W.W. for useful
discussions. This work is supported by NSF of China
(10373002, 10403001, 10273001).

\end{acknowledgements}

\end{document}